\documentclass[aps,pre,twocolumn,superscriptaddress,floatfix]{revtex4-2}

\usepackage{amsmath,amssymb,bm,mathtools}
\usepackage{graphicx,booktabs,microtype,xcolor}
\usepackage[colorlinks=true,citecolor=blue!60!black,linkcolor=blue!60!black,urlcolor=blue!60!black]{hyperref}

\newcommand{\dd}{\mathrm{d}}
\newcommand{\cE}{\mathcal E}

\newcommand{\cS}{\mathcal S}
\newcommand{\avg}[1]{\left\langle #1\right\rangle}

\makeatletter\begin{document}

\title{Particle Contacts Generate Fractional Density Relaxation}

\author{Hu Cang}
\email{cangh@uci.edu}
\affiliation{Department of Developmental and Cell Biology, University of California, Irvine, Irvine, California 92697, USA}

\date{August 6, 2026}

\begin{abstract}
Dense-liquid relaxation evolves from local particle collisions to cooperative structural rearrangements.  While hard-sphere kinetics determines an early $t^{3/2}$ fractional decay in density correlation functions, collective theories describe the subsequent structural relaxation.  A central open question has been how short-time contact physics supplies an exact starting point for the memory kernel governing later times without being modified by subsequent many-body rearrangements.  Here we resolve this problem for a broad class of reversible Brownian systems.  We prove that hard particle contacts act as reflecting boundaries in configuration space, uniquely dictating the amplitude of the leading $t^{3/2}$ density relaxation.  Mechanistically, diffusion samples a contact boundary layer of thickness $O(\sqrt{t})$, which combines with the local density response to produce the fractional signal.  We derive an explicit surface formula expressing this amplitude in terms of equilibrium contact probability, normal mobility, and density sensitivity.  For monodisperse hard spheres, this yields an exact, fit-free prediction determined entirely by static structure $S(k)$, radial contact value $g(\sigma^+)$, and short-time diffusion $D_0$.  Extending the construction, we determine the corresponding normalization for soft interfaces and prove via a Gram--Schur projection hierarchy that regular collective variables leave the leading contact amplitude strictly invariant.  The resulting formulation connects microscopic collision kinetics directly to caging and glass-like structural relaxation, providing an exact microscopic boundary condition for scattering experiments, molecular simulations, and memory-kernel reconstructions.
\end{abstract}

\maketitle

\section{Introduction}

How does a dense liquid forget its initial structure?  Scattering experiments and simulations answer this question through the intermediate scattering function $F(k,t)$, where $k$ selects the observed length scale.  Its early decay reflects diffusion and direct particle contacts; its later decay reflects coordinated many-particle rearrangement.  These regimes belong to one measured correlation, yet they are organized by two complementary theories.

The common language is an exact memory equation.  With $\phi_k(t)=F(k,t)/F(k,0)$, the normalized correlation obeys
\begin{equation}
 \begin{aligned}
 \dot{\phi}_k(t)+\Gamma_k\phi_k(t)
 &-\int_0^t\Sigma_k(t-\tau)\phi_k(\tau)\,\dd\tau=0,\\
 \phi_k(0)&=1.
 \end{aligned}
 \label{eq:memorybridge}
\end{equation}
Here $\Gamma_k$ is the instantaneous diffusive decay rate.  The kernel $\Sigma_k$ is the memory returned by particle motion that is hidden from the observed density.  Equation~\eqref{eq:memorybridge} spans the full relaxation: contact motion determines its leading short-time structure, and collective rearrangements build its later-time content.

Hard contact imposes the short-time boundary condition
\begin{equation}
 \phi_k(t)=1-\Gamma_k t+B_k t^{3/2}+O(t^2),
 \qquad t\to0,
 \label{eq:introshort}
\end{equation}
where $B_k$ is the amplitude of the first contact-induced fractional correction.  In Eq.~\eqref{eq:memorybridge}, the same physics appears as a leading $t^{-1/2}$ contact contribution to $\Sigma_k$.  Classical hard-particle calculations established this fractional signature \cite{Ackerson1976,CichockiHess1987,CichockiFelderhof1994}; Mori--Zwanzig projection and mode-coupling theory (MCT) organize the collective part of the memory \cite{Zwanzig1960,Mori1965,BengtzeliusGotzeSjolander1984,Gotze2009}.  The specific knowledge gap is to derive $B_k$ directly from general many-body contact geometry, with the correct boundary normalization, and to embed that result in the full kernel $\Sigma_k$ while preserving its short-time value.

\begin{figure*}[t]
\centering
\includegraphics[width=0.96\textwidth]{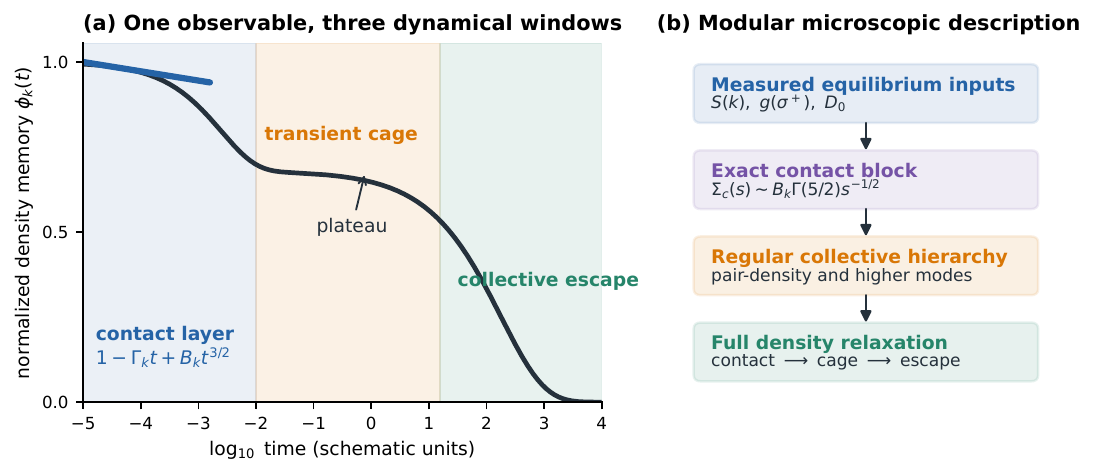}
\caption{Roadmap from microscopic contact to collective relaxation.  (a) A schematic normalized density correlator displays the short-time contact layer, a transient-cage plateau, and collective structural escape.  The blue segment highlights the local expansion derived in this article.  (b) Independently measurable equilibrium inputs determine the contact block, regular projection modes describe collective relaxation, and both contributions enter the full density correlation.}
\label{fig:roadmap}
\end{figure*}

We derive this connection for reversible overdamped dynamics with piecewise-smooth contact faces and positive normal mobility.  The resulting contact-surface formula for $B_k$ contains three inputs: equilibrium contact weight, mobility normal to the contact surface, and sensitivity of the density wave to contact motion.  For identical hard spheres, all three are fixed by measurable static and transport quantities.  A separate calculation gives the full-line normalization for a soft interface.  Gram--Schur projection then shows how regular collective modes can be added without changing the contact amplitude.  Figure~\ref{fig:roadmap} summarizes these steps from measured inputs to the complete density correlation.

\subsection{Physical origin and historical development}

The $t^{3/2}$ power can be understood by resolving diffusion normal to a contact surface.  In the space of all particle positions, each hard contact is a reflecting boundary.  During time $t$, diffusion samples a layer of thickness $O(\!\sqrt{D_c t}\,)$ next to that boundary, so the equilibrium weight involved in reflection scales as $\sqrt t$.  The normal displacement is also $O(\!\sqrt t\,)$, and its quadratic contribution to the density correlation scales as $t$.  Multiplying these two factors gives $t^{3/2}$.  This local argument fixes the exponent; its many-body amplitude requires an integral over every contact face, weighted by equilibrium structure, normal mobility, and the response of the chosen density wave.  At higher density, repeated contacts develop into localization and cooperative structural relaxation, as observed across colloidal and glass-forming systems \cite{PuseyVanMegen1986,vanMegenUnderwood1993,Angell1995,DebenedettiStillinger2001,BerthierBiroli2011}.

The short-time branch of the theory developed through Brownian kinetic calculations.  Ackerson's 1976 analysis of interacting Brownian particles identified the nonanalytic contact contribution to density correlations \cite{Ackerson1976}.  Cichocki and Hess subsequently formulated the dynamic structure factor through a memory function \cite{CichockiHess1987}.  Cichocki and Felderhof then obtained explicit short-time hard-sphere results for semidilute and dense suspensions \cite{CichockiFelderhof1993,CichockiFelderhof1994}; later work connected these corrections with colloidal diffusion and structure \cite{Banchio2000}.  These calculations established the fractional exponent for classical hard spheres and related its amplitude to contact statistics and transport.

A second branch developed through projection-operator theory.  Zwanzig introduced ensemble projection in 1960, and Mori's 1965 formulation expressed transport and collective motion through generalized Langevin equations and memory kernels \cite{Zwanzig1960,Mori1965}.  In 1984, Bengtzelius, G{\"o}tze, and Sj{\"o}lander and, independently, Leutheusser formulated the density-mode feedback mechanism that became standard MCT \cite{BengtzeliusGotzeSjolander1984,Leutheusser1984}.  Subsequent analyses organized $\beta$ relaxation and the broader liquid-glass scenario \cite{GotzeSjogren1991,GotzeSjogren1992,Gotze2009,Janssen2018}.  Kawasaki's irreducible-memory construction and Szamel's diagrammatic formulation refined how Brownian force sectors are separated and resummed \cite{Kawasaki1995,Szamel2007,Szamel2013}.

Higher-order theories extended this program beyond the standard pair-density closure.  High-order and generalized mode-coupling approaches introduced additional correlation levels \cite{WuCao2005,Mayer2006,Janssen2015}.  Recent microscopic implementations treat wave-vector-dependent hierarchies and mixtures \cite{Luo2020,Debets2021}, while direct reconstructions compare successive projected kernels with exact dynamics \cite{Pihlajamaa2023,Pihlajamaa2024}.  In the present formulation, these projected variables occupy the regular sector of the memory, while the contact calculation fixes their shared leading short-time term.

The two historical branches meet at a domain issue.  A density wave is smooth inside configuration space yet carries finite normal flux at a reflecting contact face; its generator action is therefore a surface functional.  Placing this contact sector in the Dirichlet-form domain allows the flux to pair continuously with response fields.  A weak Feshbach identity then maps the surface functional to the contact part of $\Sigma_k$, while regular collective modes supply the remaining memory.  The short-time surface term and the later projected modes thereby enter separate, explicitly defined sectors of the single kernel in Eq.~\eqref{eq:memorybridge}.

\subsection{Why the connection matters}

The contact coefficient provides an independently calibrated microscopic boundary condition.  In colloidal suspensions, particle size, static structure, contact value, and short-time diffusivity can be measured separately \cite{PuseyVanMegen1986,vanMegenUnderwood1993,vanMegenUnderwood1994}.  Broadband optical studies likewise resolve the crossover from fast localized dynamics to mode-coupling relaxation over wide time windows \cite{Cang2003JCP,Cang2003LC,Cang2005}.  Once the contact amplitude is fixed from these inputs, collective theories can devote their hierarchy to localization, cooperative rearrangement, and structural relaxation.  The same separation gives simulations a direct test across wave number before later-time closure choices enter.

Section~II identifies the contact flux in the form domain.  Section~III carries that flux into the memory and derives the many-body surface formula and its hard-sphere specialization.  Sections~IV and V establish the soft-interface calibration and the addition of regular collective modes, and Sec.~VI gives direct tests.  The derivation covers reversible Brownian hard-particle dynamics with independent or other regular mobility tensors.  Its geometric formulation also provides a reference point for extensions involving lubrication, inertia, or driven motion.

\begin{figure*}[t]
\centering
\includegraphics[width=0.96\textwidth]{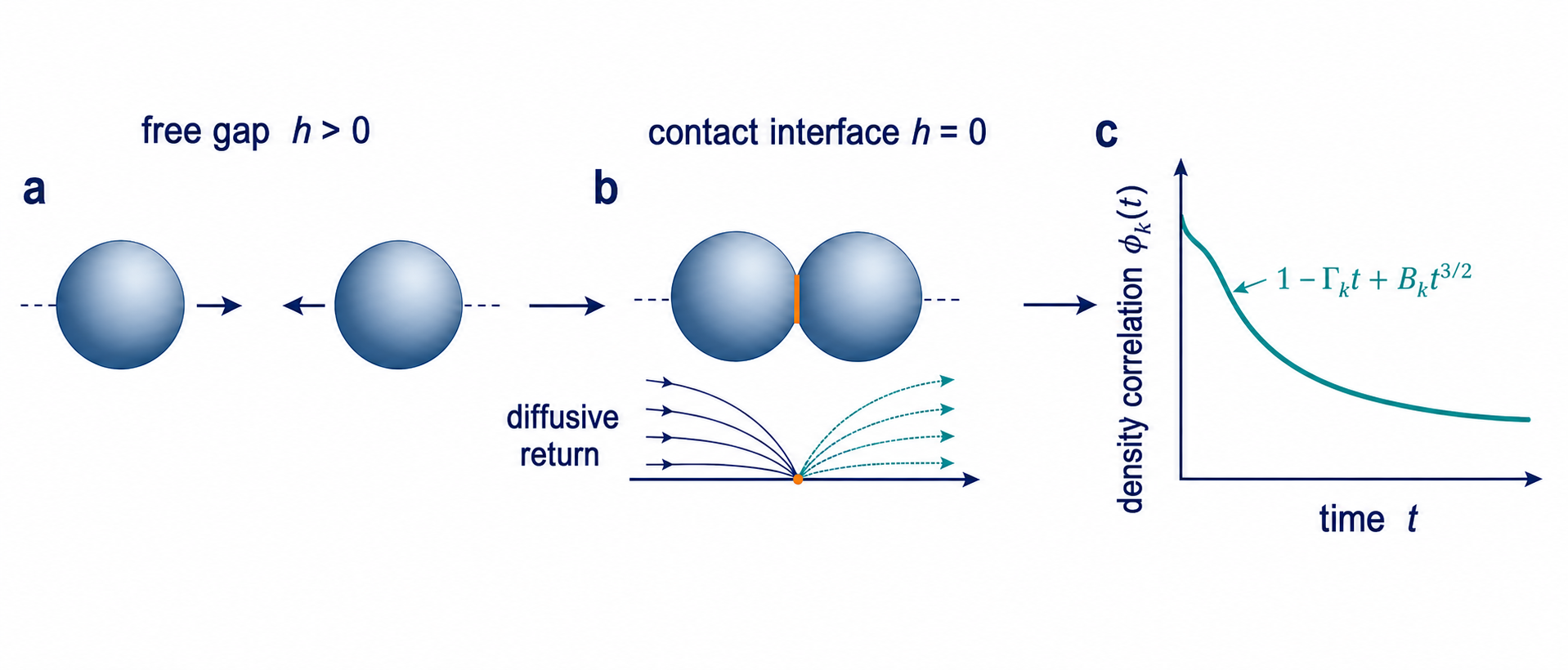}
\caption{Contact origin of the fractional density term.  (a) A pair gap $h$ approaches zero.  (b) Reflected diffusion explores a boundary layer of width $O(\!\sqrt{D_c t}\,)$ at the contact face.  (c) The normalized density correlation begins as $\phi_k(t)=1-\Gamma_k t+B_k t^{3/2}+O(t^2)$.  The curve is schematic; Eq.~\eqref{eq:Bsurface} gives the coefficient.}
\label{fig:picture}
\end{figure*}

\section{Locating contact flux in reversible diffusion}

Let $X$ denote a configuration in the allowed domain $\Omega$, with equilibrium probability measure $\mu(\dd X)$.  Let $\bm D(X)$ be a symmetric mobility--diffusion tensor that is uniformly positive in a neighborhood of every regular contact face.  Detailed balance associates the nonnegative operator $A=-\Omega^\dagger$ with the closed Dirichlet form
\begin{equation}
 \cE(f,g)=\int_\Omega (\nabla f)^*\!\cdot\bm D\nabla g\,\dd\mu,
 \qquad f,g\in D(\cE).
 \label{eq:form}
\end{equation}
The operator domain $D(A)$ imposes zero conormal flux on reflecting faces, whereas the form domain $D(\cE)$ retains every finite-energy function together with its boundary trace \cite{ArendtTerElst2011,Grebenkov2019}.

For $\rho_{\bm k}=\sum_{j=1}^N e^{i\bm k\cdot\bm r_j}$, define
\begin{equation}
 u_k=\frac{\rho_{\bm k}}{\sqrt{N S(k)}},\qquad
 \phi_k(t)=\avg{u_k^*e^{-tA}u_k},
 \label{eq:uk}
\end{equation}
 so that $\avg{|u_k|^2}=1$ and $\phi_k(t)=F(k,t)/S(k)$.  The finite first form moment is
\begin{equation}
 \Gamma_k=\cE(u_k,u_k),\qquad
 \phi_k(t)=1-\Gamma_k t+B_k t^{3/2}+O(t^2).
 \label{eq:short}
\end{equation}
For $\bm D=D_0\bm I$, one has $\Gamma_k=D_0k^2/S(k)$.

Let the regular contact faces be $\cS_c=\{X:h_c(X)=0\}$, with allowed configurations at $h_c>0$.  Introduce the local normal diffusivity and conormal density flux
\begin{equation}
 D_c=(\nabla h_c)^T\bm D\nabla h_c,\qquad
 q_{k,c}=(\nabla h_c)^T\bm D\nabla u_k.
 \label{eq:normaldata}
\end{equation}
The density mode satisfies $u_k\in D(\cE)$ and generically carries $q_{k,c}\ne0$, placing it in the finite-energy form domain.  Green's identity represents its strong generator action through a boundary functional proportional to $q_{k,c}$.  This domain structure gives a finite first derivative of $\phi_k$ and allows the next hierarchy level to be evaluated through weak pairings of the contact distribution.

\section{Carrying contact flux into density memory}

Let $P=|u_k\rangle\langle u_k|$ and $Q=1-P$.  The contact source enters the projection hierarchy as the bounded functional
\begin{equation}
 \ell_k(v)=\cE(u_k,v),\qquad v\in QD(\cE).
 \label{eq:source}
\end{equation}
Because $u_k\in D(\cE)$, Cauchy--Schwarz in the form norm makes Eq.~\eqref{eq:source} well defined when $Au_k$ contains a surface distribution.  The exact form Schur complement, which supplies the Laplace-space memory in Eq.~\eqref{eq:memorybridge}, is
\begin{equation}
 \Sigma_k(s)=\sup_{v\in QD(\cE)}
 \left\{2\,\mathrm{Re}\,\ell_k(v)-s\lVert v\rVert^2-\cE(v,v)\right\},
 \label{eq:formfeshbach}
\end{equation}
and the normalized density resolvent becomes
\begin{equation}
 \widehat\phi_k(s)=\avg{u_k^*(s+A)^{-1}u_k}
 =\frac{1}{s+\Gamma_k-\Sigma_k(s)}.
 \label{eq:densityres}
\end{equation}
For $u_k\in D(A)$, one has $\ell_k(v)=\langle QAu_k,v\rangle$, and Eq.~\eqref{eq:formfeshbach} recovers the ordinary Feshbach formula \cite{Feshbach1958,Feshbach1962}.  At hard contact, Eq.~\eqref{eq:formfeshbach} extends the same identity to the form functional generated by the surface flux.

Laplace transformation of Eq.~\eqref{eq:short} shows that
\begin{equation}
 \Sigma_k(s)=B_k\Gamma(5/2)s^{-1/2}+O(s^{-1}),
 \qquad s\to\infty.
 \label{eq:fingerprint}
\end{equation}
The local variational problem normal to a reflecting face is a half-line problem.  A boundary source $q$ produces the shifted response $|q|^2/\sqrt{D_c s}$, whereas the corresponding full-line response is $|q|^2/(2\sqrt{D_c s})$.  Summing regular faces gives
\begin{equation}
 \boxed{
 B_k^{\rm hard}=\frac{4}{3\sqrt\pi}
 \sum_c\avg{\delta(h_c)
 \frac{|q_{k,c}|^2}{\sqrt{D_c}}}_\mu .}
 \label{eq:Bsurface}
\end{equation}
Here $\avg{\delta(h_c)X}_\mu$ denotes the equilibrium contact-surface integral in the coordinate $h_c$.  Equation~\eqref{eq:Bsurface} is invariant under a positive rescaling of $h_c$: the delta, conormal flux, and $D_c$ factors compensate.  It is also nonnegative, as required by the return-memory interpretation.  Its hypotheses and half-line derivation are given in the Appendix.

For monodisperse hard spheres of diameter $\sigma$ with $\bm D=D_0\bm I$, $D_c=2D_0$ and Eq.~\eqref{eq:Bsurface} can be evaluated using the contact value $g(\sigma^+)$.  In three dimensions,
\begin{equation}
 \boxed{
 \begin{aligned}
 B_k^{\rm HS}={}&\frac{4\sqrt\pi}{9}\,
 \frac{\rho\sigma^2g(\sigma^+)(2D_0)^{3/2}k^2}{S(k)}\\
 &\times\left[1-j_0(k\sigma)+2j_2(k\sigma)\right].
 \end{aligned}}
 \label{eq:BhardSphere}
\end{equation}
The angular factor agrees with the contact factor in classical hard-sphere short-time calculations \cite{CichockiFelderhof1994,Banchio2000}.  Equation~\eqref{eq:BhardSphere} also has the required conservation limit $B_k=O(k^4)$ as $k\to0$.

\begin{figure*}[t]
\centering
\includegraphics[width=0.96\textwidth]{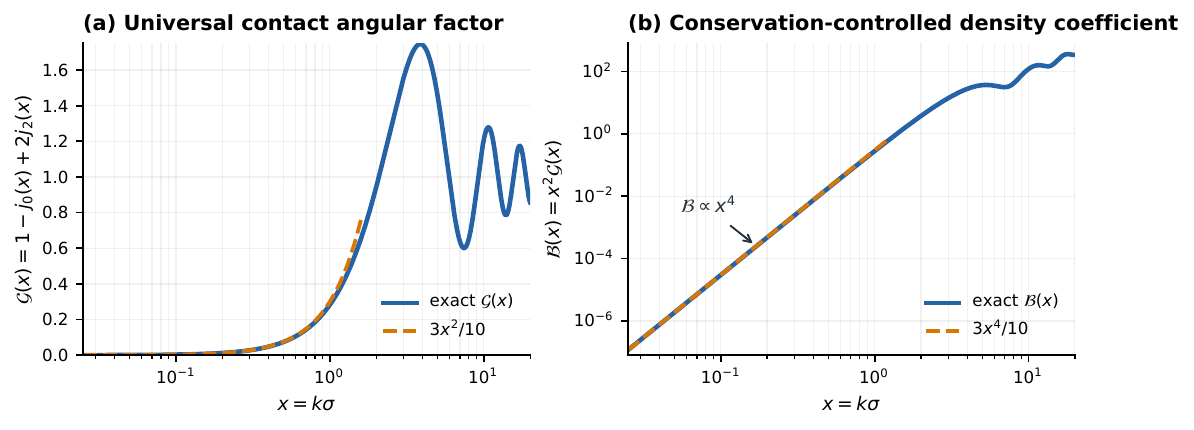}
\caption{Universal wave-number dependence of the hard-sphere contact coefficient.  (a) The angular factor $\mathcal G(x)$ from Eq.~\eqref{eq:Bscaled}; the dashed curve is its small-$x$ limit.  (b) The reduced coefficient $\mathcal B(x)=x^2\mathcal G(x)$ and its quartic conservation law.  Static structure enters the dimensional prediction through the prefactor and $S(k)$ in Eq.~\eqref{eq:BhardSphere}, while the curves shown here depend solely on $k\sigma$.}
\label{fig:hsprediction}
\end{figure*}

To isolate the universal wave-number dependence, define $x=k\sigma$ and the reduced coefficient
\begin{equation}
 \begin{aligned}
 \mathcal B(x)
 &\equiv
 \frac{9S(k)B_k^{\rm HS}}
 {4\sqrt{\pi}\rho g(\sigma^+)(2D_0)^{3/2}}\\
 &=x^2\mathcal G(x),\qquad
 \mathcal G(x)=1-j_0(x)+2j_2(x).
 \end{aligned}
 \label{eq:Bscaled}
\end{equation}
At wavelengths much larger than the particle diameter,
\begin{equation}
 \mathcal G(x)=\frac{3x^2}{10}+O(x^4),
 \qquad
 \mathcal B(x)=\frac{3x^4}{10}+O(x^6).
 \label{eq:Bsmallk}
\end{equation}
The first factor of $x^2$ comes from the density gradient at contact, and the second expresses cancellation of a uniform translation for a conserved density.  Figure~\ref{fig:hsprediction} shows how this conservation regime crosses into particle-scale oscillations once the wavelength resolves the contact shell.

\section{Calibrating soft interfaces against hard contacts}

A one-sided quadratic soft potential,
\begin{equation}
 \bar v(h)=\frac{\epsilon h^2}{2}\bm1_{h<0},
 \qquad f(h)=\epsilon h\bm1_{h<0},
 \label{eq:sqs}
\end{equation}
creates an internal interface with configurations on both sides of $h=0$.  The force is continuous, and its derivative has the jump $[f']_0=f'(0^+)-f'(0^-)=-\epsilon$.  For a general continuous observable with derivative jump $[\partial_h f]_0$, the full-line Green function gives
\begin{equation}
 B_f^{\rm int}=\frac{2}{3\sqrt\pi}
 \avg{\delta(h)D_h^{3/2}|[\partial_h f]_0|^2}.
 \label{eq:Binterface}
\end{equation}
The coefficient is one half of the reflecting-boundary result with the corresponding one-sided slope.  The two formulas therefore provide geometry-specific normalizations for soft interfaces and hard spheres.

As a calibration, consider the dimensionless bare SQS gap generator
\begin{equation}
 L_0=\partial_h^2+[1-\bar v'(h)]\partial_h
 \label{eq:L0}
\end{equation}
in the weighted space $\dd\mu_\epsilon=e^{h-\bar v(h)}\dd h$, and define
\begin{equation}
 M_{\rm bare}(t)=\frac{\widehat\varphi}{2}
 \int f(h)[e^{tL_0}f](h)\,\dd\mu_\epsilon(h).
 \label{eq:Mbare}
\end{equation}
Here $\widehat\varphi$ denotes the scaled packing fraction used in the infinite-dimensional gap theory \cite{Maimbourg2016,Kurchan2016,Manacorda2020}.  Equation~\eqref{eq:Binterface} yields
\begin{equation}
 M_{\rm bare}(t)=M_0+M_1t+
 \frac{\widehat\varphi\epsilon^2}{3\sqrt\pi}t^{3/2}+O(t^2).
 \label{eq:sqscoef}
\end{equation}
The exact one-gap resolvent also produces $t^{5/2},t^{7/2},\ldots$, interlaced with integer powers.  Equation~\eqref{eq:Binterface} fixes the leading local interface coefficient, while the subleading coefficients characterize this fully specified one-gap model.

\section{Building collective memory around contact}

At a fixed shift $s$, let $r_c$ be the exact Riesz vector of the contact functional in the shifted form metric
\begin{equation}
 \langle f,g\rangle_s=s\langle f,g\rangle+\cE(f,g).
\end{equation}
Add regular vectors $r_1,\ldots,r_m$ and form their Gram matrix $G_s$.  The overlap vector of the \emph{pure contact functional} is
\begin{equation}
 (B_c)_i=\langle r_i,r_c\rangle_s=(G_s)_{i0}.
\end{equation}
Consequently,
\begin{equation}
 G_s^{-1}B_c=\bm e_0,
 \qquad B_c^\dagger G_s^{-1}B_c=(G_s)_{00}.
 \label{eq:gramlock}
\end{equation}
This exact first-column identity locks the pure contact block.  General physical sources are organized by writing
\begin{equation}
G_s=\begin{pmatrix}\Sigma_c&c^\dagger\\c&G_R\end{pmatrix}.
\end{equation}
The block elimination yields the regular Schur complement $G_R-cc^\dagger/\Sigma_c$ and the correspondingly contact-subtracted regular source; the explicit formula is given in the Appendix.  Added modes with bounded contact traces and $L^2$ sources have self and mixed responses of $O(s^{-1})$, preserving the $s^{-1/2}$ contact coefficient while enriching finite-frequency and subleading terms.  A mode carrying a surface distribution joins the contact sector and contributes through the same interfacial analysis.

\section{Testing the fractional coefficient}

\begin{figure*}[t]
\centering
\includegraphics[width=0.92\textwidth]{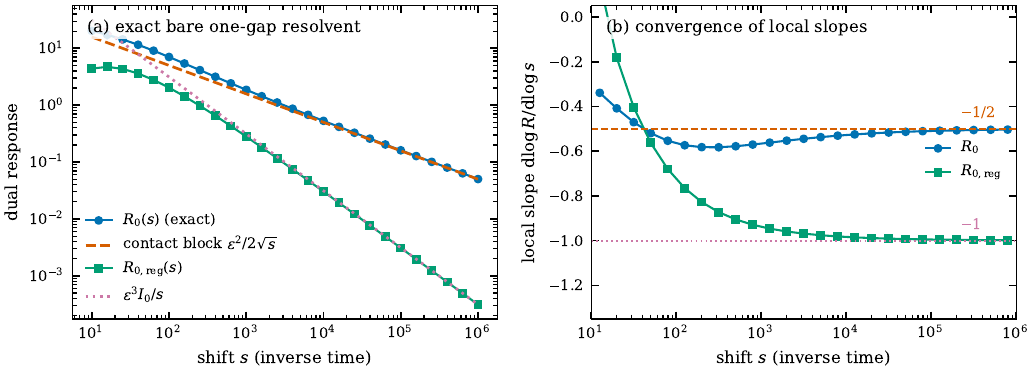}
\caption{Exact bare SQS one-gap resolvent for $\epsilon=10$, evaluated entirely from the stated expressions.  (a) The dual response $R_0$, the predicted internal-interface block $\epsilon^2/(2\sqrt{s})$, and the residual approaching $\epsilon^3I_0/s$.  (b) Local log slopes approach $-1/2$ before subtraction and $-1$ afterward.  Values use 80-digit parabolic-cylinder evaluation; the displayed slope is a convergence diagnostic for the analytic expansion \eqref{eq:dualexp}.}
\label{fig:gate}
\end{figure*}

For the SQS model define $C_\epsilon(s)=\langle f,(s-L_0)^{-1}f\rangle_{\mu_\epsilon}$ and
\begin{equation}
 R_0(s)=s^2\left[C_\epsilon(s)-\frac{C_0}{s}-\frac{C_1}{s^2}\right].
 \label{eq:dual}
\end{equation}
The exact one-gap expansion gives
\begin{equation}
 R_0(s)=\frac{\epsilon^2}{2\sqrt{s}}
 +\frac{\epsilon^3 I_0(\epsilon)}{s}+O(s^{-3/2}),
 \label{eq:dualexp}
\end{equation}
where
\begin{equation}
 I_0(\epsilon)=e^{1/(2\epsilon)}
 \sqrt{\frac{\pi}{2\epsilon}}\,
 \operatorname{erfc}\!\left(\frac{1}{\sqrt{2\epsilon}}\right).
\end{equation}
After the full-line contact block is subtracted, the residual is $O(s^{-1})$.  The finite dyadic slope in Fig.~\ref{fig:gate} shows its convergence to this analytically derived scaling.

For hard spheres, both the exponent and amplitude can be tested without fitting the short-time term.  Measure $S(k)$, $g(\sigma^+)$, and the short-time diffusivity independently; Eq.~\eqref{eq:BhardSphere} then fixes $B_k$.  Plotting $[\phi_k(t)-1+\Gamma_k t]/t^{3/2}$ should approach that value over the Brownian short-time window.  Scalar independent diffusivity selects Eq.~\eqref{eq:BhardSphere}, while a nonscalar positive mobility selects the general surface expression \eqref{eq:Bsurface}.

\section{Discussion}

Equation~\eqref{eq:Bsurface} identifies the fractional coefficient with an equilibrium average over contact surfaces.  Its $t^{3/2}$ power comes from two local factors: diffusion samples a layer of width $O(\!\sqrt{D_c t}\,)$, and reflection changes the density correlation by $O(t)$.  The amplitude retains the many-body information that this scaling argument cannot supply.  The contact measure gives the equilibrium weight of configurations at an encounter, $q_{k,c}$ measures the response of the observed density wave to normal motion, and $D_c$ sets the rate at which the contact layer is sampled.  For identical hard spheres, these factors reduce to the measurable quantities $S(k)$, $g(\sigma^+)$, $D_0$, and $\sigma$ in Eq.~\eqref{eq:BhardSphere}.

The factor of two between a reflecting hard contact and a soft internal interface follows from their local Green functions: diffusion occupies a half-line in the first case and a full line in the second.  The exactly solvable SQS gap verifies this normalization and shows where the leading interface term appears within the sequence of integer and half-integer powers.  Interface geometry fixes the first fractional coefficient; smooth drift, curvature, and the rest of the generator enter at subsequent orders.

The projection calculation addresses a different question: whether collective variables can alter this leading coefficient.  Equation~\eqref{eq:gramlock} fixes the pure contact response through the first column of the shifted Gram matrix.  Regular modes with bounded traces contribute at $O(s^{-1})$ and therefore leave the $s^{-1/2}$ contact term unchanged.  A mode that carries its own surface distribution is instead assigned to the contact sector.  These criteria make the separation between contact and regular memory explicit at each level of a projection hierarchy.

The resulting experimental and simulation test uses independently determined inputs.  Static structure and short-time mobility fix $\Gamma_k$ and $B_k$, after which $[\phi_k(t)-1+\Gamma_k t]/t^{3/2}$ approaches a predicted contact plateau across wave number.  Calibrating this term isolates the later memory associated with caging and structural escape.  Its wave-number dependence further separates local contact geometry from collective relaxation because Eq.~\eqref{eq:BhardSphere} fixes the contact contribution before a later-time closure is chosen.

The derivation assumes reversible overdamped motion, piecewise-smooth contact faces, and positive normal mobility.  Other short-time physics changes the local normal problem in a controlled way: lubrication modifies the mobility near contact, inertia inserts a ballistic regime, and external driving changes the stationary measure.  In each case the appropriate interfacial Green function can be matched to the regular projection sector.

MCT, generalized projection theories, and direct memory-kernel reconstructions may therefore use $\Gamma_k$ and $B_k$ as independently fixed microscopic inputs.  Contact geometry determines the initial fractional return, while projected collective modes describe cage formation and structural relaxation.  Their placement in the same memory equation connects the classical short-time contact signature to systematically refined descriptions of dense-liquid dynamics.

\begin{acknowledgments}
This work was supported by the National Human Genome Research Institute of the National Institutes of Health under Award No.~1R01HG014004-01.
\end{acknowledgments}

\section*{Data Availability}
All results in this article are analytical.

\appendix
\section{Matching time and resolvent coefficients}

Suppose
\begin{equation}
 \phi(t)=1-\Gamma t+B t^{3/2}+O(t^2).
\end{equation}
Termwise Laplace transformation gives
\begin{equation}
 \widehat\phi(s)=\frac1s-\frac{\Gamma}{s^2}
 +\frac{B\Gamma(5/2)}{s^{5/2}}+O(s^{-3}).
 \label{app:laplace}
\end{equation}
If $\Sigma(s)=\sigma_{1/2}s^{-1/2}+O(s^{-1})$, expansion of Eq.~\eqref{eq:densityres} gives
\begin{equation}
 \frac{1}{s+\Gamma-\Sigma(s)}
 =\frac1s-\frac{\Gamma}{s^2}
 +\frac{\sigma_{1/2}}{s^{5/2}}+O(s^{-3}).
\end{equation}
Therefore $\sigma_{1/2}=B\Gamma(5/2)$, which fixes all factors in Eq.~\eqref{eq:fingerprint}.

\section{Deriving the weak Schur complement}

Let $A\geq0$ be the self-adjoint operator associated with the closed form \eqref{eq:form}.  For $v,w\in QD(\cE)$ define
\begin{equation}
 \langle v,w\rangle_s=s\langle v,w\rangle+\cE(v,w).
\end{equation}
The functional $\ell_k(w)=\cE(u_k,w)$ is bounded in this norm.  Its Riesz vector $r_s\in QD(\cE)$ is uniquely specified by
\begin{equation}
 \langle r_s,w\rangle_s=\ell_k(w).
 \label{app:riesz}
\end{equation}
Completing the square yields
\begin{align}
 &2\operatorname{Re}\ell_k(v)-\lVert v\rVert_s^2\notag\\
 &\hspace{8mm}=\lVert r_s\rVert_s^2-\lVert v-r_s\rVert_s^2,
\end{align}
so Eq.~\eqref{eq:formfeshbach} equals
\begin{equation}
 \Sigma_k(s)=\lVert r_s\rVert_s^2=\ell_k(r_s).
 \label{app:sigmariesz}
\end{equation}

For completeness, let $z=(s+A)^{-1}u_k=\alpha u_k+v$, where $v\perp u_k$.  The weak resolvent equation tested against $w\in QD(\cE)$ gives $v=-\alpha r_s$.  Testing against $u_k$ then gives
\begin{equation}
 \alpha\,[s+\Gamma_k-\ell_k(r_s)]=1.
\end{equation}
Since $\alpha=\langle u_k,(s+A)^{-1}u_k\rangle$, Eqs.~\eqref{eq:densityres} and \eqref{eq:formfeshbach} follow for every $u_k\in D(\cE)$, including the contact case.

\section{Extracting the density coefficient from a reflecting face}

Choose a regular contact face and local coordinates $(h,y)$, where $h\geq0$ is the allowed normal coordinate and $y$ is tangential.  Freeze the coefficients over the $h=O(s^{-1/2})$ layer.  After choosing tangential coordinates that are orthogonal in the diffusion metric, the leading shifted energy at fixed $y$ is
\begin{equation}
 \int_0^\infty\left[s|v|^2+D_c|\partial_h v|^2\right]\dd h,
\end{equation}
and the boundary part of the source is $q_{k,c}^*v(0)$.  The maximizing field is proportional to
\begin{equation}
 v_s(h)=\frac{q_{k,c}}{\sqrt{D_c s}}
 e^{-\sqrt{s/D_c}\,h}.
\end{equation}
Substitution gives the response
\begin{equation}
 \Sigma_{k,c}(s)=\avg{\delta(h_c)
 \frac{|q_{k,c}|^2}{\sqrt{D_c}}}_\mu s^{-1/2}
 +O(s^{-1}).
 \label{app:halfspace}
\end{equation}
Tangential gradients, smooth drift, curvature, and variation of $\bm D$ enter one order later.  Under the stated piecewise-smoothness assumption, regular contact faces supply the $s^{-1/2}$ surface term, while their codimension-two intersections enter at subsequent orders.  Combining Eq.~\eqref{app:halfspace} with Appendix A gives Eq.~\eqref{eq:Bsurface} because $1/\Gamma(5/2)=4/(3\sqrt\pi)$.

We now specialize the result to identical spheres.  For $h_{ij}=|\bm r_j-\bm r_i|-\sigma$ and $\widehat{\bm n}_{ij}=(\bm r_j-\bm r_i)/|\bm r_j-\bm r_i|$, independent particle diffusivity gives
\begin{equation}
 D_{ij}=2D_0,\qquad
 \partial_{h_{ij}}\rho_{\bm k}
 =\frac{i\bm k\cdot\widehat{\bm n}_{ij}}{2}
 \left(e^{i\bm k\cdot\bm r_j}-e^{i\bm k\cdot\bm r_i}\right).
 \label{app:hsderivative}
\end{equation}
The equilibrium pair identity is
\begin{equation}
 \sum_{i<j}\avg{\delta(h_{ij})F(\widehat{\bm n}_{ij})}
 =\frac{N\rho\sigma^2g(\sigma^+)}{2}
 \int\dd\Omega_{\bm n}\,F(\bm n).
 \label{app:pairidentity}
\end{equation}
Using $u_k=\rho_{\bm k}/\sqrt{NS(k)}$, the required angular integral is
\begin{align}
 &\int\dd\Omega_{\bm n}(\bm k\cdot\bm n)^2
 [1-\cos(k\sigma\,\widehat{\bm k}\cdot\bm n)]\notag\\
 &\hspace{12mm}=\frac{4\pi k^2}{3}
 [1-j_0(k\sigma)+2j_2(k\sigma)].
\end{align}
Substitution into Eq.~\eqref{eq:Bsurface} gives Eq.~\eqref{eq:BhardSphere}.  Since $1-j_0(x)+2j_2(x)=3x^2/10+O(x^4)$, density conservation implies $B_k=O(k^4)$.

\section{Calibrating the full-line interface with the SQS model}

For an internal interface, both $h<0$ and $h>0$ are allowed.  The decaying Green function of the frozen normal operator is
\begin{equation}
 (s-D_h\partial_h^2)G_s(h)=\delta(h),\qquad
 G_s(h)=\frac{e^{-\sqrt{s/D_h}|h|}}{2\sqrt{D_h s}}.
 \label{app:fullgreen}
\end{equation}
A derivative jump $a=[\partial_h f]_0$ produces the distributional source $D_h a\delta(h)$.  Its self-response is therefore $D_h^{3/2}|a|^2s^{-1/2}/2$.  Division by $\Gamma(5/2)$ gives Eq.~\eqref{eq:Binterface}.  The factor $1/2$ in Eq.~\eqref{app:fullgreen} is absent for a single reflecting half-line, which proves the factor-of-two distinction used in the main text.

The bare SQS example can be solved exactly.  For $h>0$, the homogeneous solution of $(s-L_0)\psi=0$ is $e^{r_-h}$ with
\begin{equation}
 r_-=\frac{-1-\sqrt{1+4s}}{2}.
\end{equation}
For $h<0$, set $z=\sqrt\epsilon h-1/\sqrt\epsilon$.  A homogeneous solution is
\begin{equation}
 e^{z^2/4}D_{-s/\epsilon}(-z),
\end{equation}
where $D_\nu$ is a parabolic-cylinder function.  Adding the affine particular solution for the source $f(h)=\epsilon h$ and matching the value and derivative at $h=0$ determines $C_\epsilon(s)$ uniquely.

Writing
\begin{equation}
 I_0(\epsilon)=\int_{-\infty}^0
 e^{h-\epsilon h^2/2}\dd h,
\end{equation}
the resulting time expansion of Eq.~\eqref{eq:Mbare} through order $t^{5/2}$ is
\begin{align}
 M_0&=\frac{\widehat\varphi}{2}[(1+\epsilon)I_0-1],\\
 M_1&=-\frac{\widehat\varphi\epsilon^2I_0}{2},\\
 M_{3/2}&=\frac{\widehat\varphi\epsilon^2}{3\sqrt\pi},\\
 M_2&=\frac{\widehat\varphi\epsilon^3I_0}{4},\\
 M_{5/2}&=-\frac{\widehat\varphi\epsilon^2(15\epsilon+1)}{60\sqrt\pi}.
 \label{app:sqscoeffs}
\end{align}
The last two values are bare one-gap coefficients.  Retarded friction or a self-consistent memory can change them; Eq.~\eqref{eq:sqscoef} is the local interface coefficient.

For the unweighted correlation $C_\epsilon(t)=2M_{\rm bare}(t)/\widehat\varphi$, Laplace transformation of Eq.~\eqref{app:sqscoeffs} gives Eqs.~\eqref{eq:dual} and \eqref{eq:dualexp}.  In particular, the $t^{3/2}$ term becomes $\epsilon^2s^{-5/2}/2$, while the $t^2$ term becomes $\epsilon^3I_0s^{-3}$.  Multiplication by $s^2$ yields the two displayed powers in Eq.~\eqref{eq:dualexp}.

\section{Separating contact and regular sources}

Let $r_c,r_1,\ldots,r_m$ be linearly independent in the shifted form metric and let
\begin{equation}
 (G_s)_{ij}=\langle r_i,r_j\rangle_s.
\end{equation}
The Riesz property gives $(B_c)_i=\ell_c(r_i)=\langle r_c,r_i\rangle_s$, so $B_c$ is the first column of $G_s$.  This proves Eq.~\eqref{eq:gramlock}.  If the vectors are redundant, the same statement holds after restriction to the Gram range.

For a general additional source $d=(\rho,r)^T$ and
\begin{equation}
 G_s=\begin{pmatrix}\Sigma_c&c^\dagger\\c&G_R\end{pmatrix},
 \qquad S_R=G_R-\frac{cc^\dagger}{\Sigma_c},
\end{equation}
the block inverse gives
\begin{equation}
 d^\dagger G_s^{-1}d=
 \frac{|\rho|^2}{\Sigma_c}
 +\left(r-\frac{c\rho}{\Sigma_c}\right)^\dagger
 S_R^{-1}
 \left(r-\frac{c\rho}{\Sigma_c}\right).
 \label{app:mixedschur}
\end{equation}
Thus the regular coordinates are the contact-subtracted combinations displayed in Eq.~\eqref{app:mixedschur}.  For an $L^2$ regular source, the self-response is bounded by $\lVert r\rVert^2/s$; a bounded trace also makes its pairing with the exponentially localized contact Riesz field $O(s^{-1})$.  Under these explicit conditions the contact block carries the $s^{-1/2}$ coefficient.  A mode with its own surface distribution belongs to the contact sector and receives the corresponding interfacial treatment.

\section{Reproducing the exact resolvent test}

Figure~\ref{fig:gate} is evaluated directly from the exact one-gap expressions at shifts $10\leq s\leq10^6$ and $\epsilon=10$, using 80-digit internal arithmetic.  The resolvent values agree with an independent finite-volume discretization.

\end{document}